\documentclass[pra,twocolumn,amsmath,amssymb,aps,superscriptaddress]{revtex4}
\usepackage{amssymb, amsmath,latexsym}
\usepackage[dvips]{graphicx}

\def\real{\mathbb{R}}

\def\integer{\mathbb{Z}}

\def\Tr{\mathop{\rm Tr}\nolimits}

\def\Label#1{\label{#1}\ [\ #1\ ]\ }
\def\Label{\label}

\begin{document}

\title{Phase estimation with photon number constraint}

\author{Masahito Hayashi}
\affiliation{Graduate School of Information Sciences, Tohoku University, Sendai, 980-8579, Japan}
\affiliation{Centre for Quantum Technologies, National University of Singapore, 3 Science Drive 2, Singapore 117542}


\begin{abstract}
Many researches 
proposed the use of the noon state as the input state for phase estimation, which is one topic of quantum metrology.
This is because the input noon state provides the maximum Fisher information at the specific point.
However, the Fisher information does not necessarily give the attainable bound for estimation error.
In this paper, we adopt the local asymptotic mini-max criterion as well as the mini-max criterion,
and show that
the maximum Fisher information does not give the attainable bound for estimation error under
these criteria in the phase estimation.
We also propose the optimal input state under the constraints for photon number of the input state
instead of the noon state.
\end{abstract}

\maketitle

\section{Introduction}\Label{s1}
As one of the most simple methods in quantum metrology,
phase estimation has been treated by many authors\cite{Luis,BDR,IH,Kitaev,Giovannetti1,Giovannetti2} and 
its experimental demonstration has been reported by many groups\cite{Heisenberg-limit,take1,take2,Mag}.
As is pointed by \cite{PSKHB}, the phase estimation with Mach-Zehnder (MZ) interferometer is important
from the view point of the standard configuration.
Many experiment groups realized the noon state.
The reason is why the noon state provides the optimal Fisher information.
In the independent and identical distributed setting of state estimation, 
Fisher information provides the asymptotic bound of mean square error (MSE)\cite{4,5,F1}.
However, the phase estimation does not fall in this setting.
Hence, it has not been discussed sufficiently whether the minimum estimation error is realized when the noon state is inputed.

In this paper, we treat the phase estimation with two kinds of photon number constraints.
One is 
the constraint for the average of the square of the photon number concerning the input state.
The other is 
the constraint for the maximum photon number concerning the input state.
In order to treat the phase estimation, we focus on two criteria.
One is the minimum mean square error (MSE) under the locally unbiased (LUB) condition,
which coincides with the SLD Fisher information \cite{1,2,3}.
The other is the mini-max value of the MSE, which minimizes the worst MSE.
The latter is essentially equivalent with the optimization with the covariant restriction for measurement\cite{HoC}.
As an intermediate concept, we focus on the local asymptotic mini-max criterion,
which was introduced by Hajek \cite{Hajek} and was applied to quantum channel estimation by Hayashi \cite{Ha5}.
Then, we can define three kinds of bounds under two kinds of constraint.
The main purpose of this paper is the comparison
between these three bounds with both constraints.
The second purpose is to check the validity of use of the noon state as the input state
and seeking the optimal input state.

The organization of this paper is the following.
In section \ref{s2}, we treat the phase estimation problem under the locally unbiased condition.
In section \ref{s3}, we clarify the relation between the single-application case and the multiple-application case.
In section \ref{s4}, 
we discuss the locally asymptotic mini-max criterion.
In section \ref{s5}, 
we treat the global mini-max criterion and the relation among three kinds of criteria under the asymptotic limit.
In section \ref{s6}, 
we treat the MSE of the noon state with both of the global and local asymptotic mini-max criteria.

\section{SLD Fisher information and locally unbiased condition}\Label{s2}
In the two-mode photonic system
$L^2(\real^2)$, 
the phase operator $U_\theta$ is given by 
\begin{align*}
U_\theta:=
\sum_{n,m\ge 0} e^{i(n-m)\theta}|n,m\rangle \langle n,m|
\end{align*}
with $\theta \in [-\pi,\pi]$.
When the true parameter $\theta$ is unknown, 
we can estimate the parameter by inputing the known state $|\phi\rangle$
and measuring the output state $U_\theta |\phi\rangle$.
Hence, when the measurement corresponds to the POVM $M=\{M_{\hat{\theta}}\}$,
our estimator is described by the pair of the input state $|\phi\rangle$ and the POVM $M$ as Fig \ref{f1}.

\begin{figure}[htbp]
\begin{center}
\scalebox{1.0}{\includegraphics[scale=0.3]{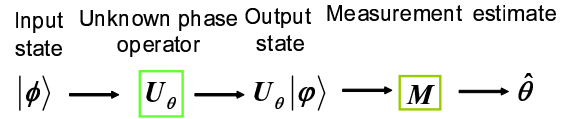}}
\end{center}
\caption{Scheme of phase estimation}
\Label{f1}
\end{figure}%

In this case, the estimate $\hat{\theta}$ obeys the distribution 
$\langle \phi| U_\theta^\dagger M_{\hat{\theta}} U_\theta |\phi\rangle$.
The mean square error (MSE) is given by
\begin{align*}
E_\theta (|\phi\rangle,M):=
\int_{-\pi}^{\pi}
(\theta-T_\theta(\hat{\theta}))^2
\langle \phi| U_\theta^\dagger M_{\hat{\theta}} U_\theta |\phi\rangle
d \hat{\theta},
\end{align*}
where
\begin{align*}
T_\theta(\hat{\theta}):=
\left\{
\begin{array}{ll}
\hat{\theta} +2 \pi & \hbox{ if }\hat{\theta}  < \theta- \pi \\
\hat{\theta} & \hbox{ if } \theta- \pi\le \hat{\theta}  < \theta+ \pi \\
\hat{\theta} -2 \pi & \hbox{ if } \theta+ \pi \le \hat{\theta}  .
\end{array}
\right.
\end{align*}
When the input state $U_\theta |\phi\rangle$ is given, 
our problem is estimation of the parameter $\theta$ with the pure state family
$\{U_\theta |\phi\rangle \langle \phi| U_\theta^\dagger\}$.
In order to treat the lower bound of the MSE, 
we focus on the symmetric logarithmic derivative (SLD) Fisher information\cite{FN}
\begin{align}
J_\theta(|\phi\rangle)=
4\langle \phi| \frac{d U_\theta}{d \theta}^\dagger  \frac{d U_\theta}{d \theta} |\phi\rangle
-4 |\langle \phi| \frac{d U_\theta}{d \theta} |\phi\rangle|^2.
\Label{12-29-3}
\end{align}
This equation can be shown by the following way:
$\frac{d}{d\theta}U_\theta |\phi\rangle \langle \phi| U_\theta^\dagger
=X |\phi\rangle \langle \phi|+ |\phi\rangle \langle \phi|X^\dagger$,
where $X= \frac{d U_\theta}{d \theta}- \langle \phi| \frac{d U_\theta}{d \theta} |\phi\rangle$.
Thus,
$
\frac{d}{d\theta}U_\theta |\phi\rangle \langle \phi| U_\theta^\dagger
=2 (X |\phi\rangle \langle \phi|+ |\phi\rangle \langle \phi|X^\dagger )\circ |\phi\rangle \langle \phi|$,
where $A\circ B= \frac{AB+BA}{2}$.
Then, $J_\theta(|\phi\rangle)=
\Tr (2 (X |\phi\rangle \langle \phi|+ |\phi\rangle \langle \phi|X^\dagger ))^2 |\phi\rangle \langle \phi|$,
which equals the right hand side of (\ref{12-29-3}).

In statistical inference, an unbiased estimator plays an important role.
However, since the parameter space is the interval $[-\pi,\pi]$,
it is difficult to define the unbiased condition.
It is possible to define the locally unbiased condition.
When the estimator $(|\phi\rangle,M)$
satisfies the condition
\begin{align*}
\int_{-\pi}^{\pi}
T_\theta(\hat{\theta})
\langle \phi| U_{\theta_0}^\dagger M_{\hat{\theta}} U_{\theta_0} |\phi\rangle
d \hat{\theta} & =\theta_0 \\
\int_{-\pi}^{\pi}
T_\theta(\hat{\theta})
\frac{d}{d\theta}
\langle \phi| U_\theta^\dagger M_{\hat{\theta}} U_\theta |\phi\rangle
|_{\theta=\theta_0}
d \hat{\theta}
& =1,
\end{align*}
it is called locally unbiased at $\theta_0$.
When an estimator $(|\phi\rangle,M)$ is locally unbiased (LUB) at $\theta_0$,
the SLD Cramer-Rao inequality
\begin{align}
E_\theta (|\phi\rangle,M)
\ge 
J_\theta(|\phi\rangle)^{-1}.\Label{10-26-1}
\end{align}
Its proof is similar to the usual case of the SLD Cramer-Rao inequality\cite{1,2,3}
because the Schwarz inequality plays the same rule.

In this problem, when we input the state with a larger photon number,
the MSE becomes smaller.
So, it is suitable to constraint the photon number of the input state.
As the first kind of constraint, we restrict the average 
of the square photon number 
${(N^2)}_{av}(|\phi\rangle):=\langle \phi| \hat{N}^2 |\phi\rangle$ of the input state $|\phi\rangle$, where
$\hat{N}:=\sum_{n,m}(n+m)|n,m\rangle \langle n,m|$.
As the second kind of constraint, we restrict the maximum  
of the photon number 
$N_{max}(|\phi\rangle):=\max_{n,m}\{n+m| \langle n,m|\phi\rangle \neq 0  \}$ of the input state $|\phi\rangle$.
So, we define the two bounds of MSE as follows:
\begin{align*}
& \tilde{C}_{av,\theta}(E) \\
:=&
\inf 
\left\{
E_\theta (|\phi\rangle,M)
\left|
\begin{array}{l}
 (|\phi\rangle,M) \hbox{is LUB at }\theta_0 \\
{(N^2)}_{av}(|\phi\rangle) \le E^2
\end{array}
\right. \right\} \\
&\tilde{C}_{max,\theta}(E) \\
:=&
\inf 
\left\{
E_\theta (|\phi\rangle,M)
\left|
\begin{array}{l}
 (|\phi\rangle,M) \hbox{is LUB at }\theta_0 \\
N_{max}(|\phi\rangle) \le E 
\end{array}
\right. \right\} .
\end{align*}
Since 
$J_\theta(|\phi\rangle)
\le 4 \langle \phi| \hat{N}^2 |\phi\rangle$,
the SLD Cramer-Rao inequality (\ref{10-26-1}) implies
\begin{align*}
\tilde{C}_{max,\theta} (E)\ge 
\tilde{C}_{av,\theta}  (E) \ge \frac{1}{4E^2} .
\end{align*}
When $E$ is an integer $n$ and 
the input state $|\phi\rangle$ is 
the noon state $|\phi_{n,noon}\rangle :=\frac{1}{\sqrt{2}}(|n,0\rangle + |0,n\rangle)$,
$J_\theta(|\phi_{n,noon}\rangle)^{-1}= \frac{1}{4n^2} $.
There exists a POVM satisfying the locally unbiased condition
whose MSE attains the inverse of the SLD Fisher information.
Hence,
\begin{align*}
\tilde{C}_{max,\theta} (n)= \tilde{C}_{av,\theta}  (n) = \frac{1}{4 n^2} .
\end{align*}

In the state estimation, it is known that 
there exists a sequence of estimators that asymptotically attains 
the inverse of the SLD Fisher information at all points\cite{4,5,F1}.
Hence, we expect that
the bounds
\begin{align}
\tilde{C}_{max,\theta} &:= \lim_{E \to \infty} E^2 \tilde{C}_{max,\theta} (E) = \frac{1}{4}
\Label{12-28-6}\\
\tilde{C}_{av,\theta} &:=\lim_{E \to \infty} E^2 \tilde{C}_{av,\theta}  (E) = \frac{1}{4} 
\Label{12-28-7}
\end{align}
can be attained asymptotically in all points by a single sequence of estimator under the respective restrictions.
However, 
the same argument does not necessarily hold in the case of channel estimation.

\section{Multiple application}\Label{s3}
As an extension of the formulation given in Section \ref{s2},
we assume that 
the $m$-fold tensor product system of $L^2(\real^2)$, i.e., $L^2(\real^2)^{\otimes m}$,
is allowed as the input system
and the unknown operator $U_\theta^{\otimes m}$ is applied.
It is required to estimate the unknown parameter $\theta$.
In this case, 
we consider the number operator $\hat{N}_m:=
\sum_{k=1}^m  I^{\otimes k-1} \otimes \hat{N}\otimes I^{\otimes m-k}$,
and the estimator is given by the pair of the input state $|\phi\rangle$ on $L^2(\real^2)^{\otimes m}$
and a POVM $M$ on $L^2(\real^2)^{\otimes m}$.
Now, we define the projection
\begin{align*}
P_{a,b}:=\sum_{\#1}\sum_{\#2}
|a_1,b_1 \rangle \langle a_1,b_1|\otimes \cdots \otimes |a_m,b_m \rangle \langle a_m,b_m|,
\end{align*}
where $\#1$ is $a_1,\ldots, a_m:\sum_j a_j=a$
and $\#2$ is $b_1,\ldots, b_m:\sum_j b_j=b$.
Then,
we define the normalized vector 
$|\phi_{a,b}\rangle := \frac{1}{\|P_{a,b}|\phi\rangle\|}P_{a,b}|\phi\rangle$
and the isometry $V_m:=\sum_{a,b} |a,b\rangle\langle \phi_{a,b}|$ from 
$L^2(\real^2)^{\otimes m}$ to $L^2(\real^2)$.
Then, the pair of the state $V_m |\phi\rangle $
and the POVM $\{ V_m M_{\hat{\theta}}V_m^\dagger \}$
gives an estimator on the single input system $L^2(\real^2)$.
The outcome of this estimator has the same statistical behavior of the original estimator 
$(|\phi\rangle, M)$ on $L^2(\real^2)^{\otimes m}$.
Further, 
the average of the square number of $V_m |\phi\rangle $ equals that of $|\phi\rangle$.
Conversely, any estimator $(|\phi\rangle ,\{M_{\hat{\theta}}\})$ on the single input system $L^2(\real^2)$
can be simulated by an estimator $(W_m^\dagger |\phi\rangle ,\{W_m^\dagger M_{\hat{\theta}}V_m\})$ on $L^2(\real^2)^{\otimes m}$,
where $W_{m}:=\sum_{a,b} |\psi_{a,b}\rangle\langle a,b|  $
and $|\psi_{a,b}\rangle$ is a normalized vector on the range of $P_{a,b}$.
So, the optimization of estimator $(|\phi\rangle ,\{M_{\hat{\theta}}\})$ on the single input system $L^2(\real^2)$ 
with constraint concerning average of square of photon number
is mathematically equivalent with that on the $m$-fold tensor product system $L^2(\real^2)^{\otimes m}$
with constraint concerning average of square of photon number.

Next, we consider the case when only state in the subspace spanned by $\{|00\rangle,|10\rangle,|01\rangle\}$
is allowed as the input system.
In this case, the input system is 3-dimensional and is written by ${\cal H}$.
Its $m$-fold input system is given as ${\cal H}^{\otimes m}$.
For a given input state $|\phi\rangle \in {\cal H}^{\otimes m}$,
the state $V_m|\phi\rangle$ satisfies 
the condition $\max_{a,b}\{a+b| \langle a,b|V_m|\phi\rangle \neq 0  \} \le m$.
Conversely, when the input state $|\phi\rangle$ satisfies 
the condition $\max_{a,b}\{a+b| \langle a,b|\phi\rangle \neq 0  \} \le m$,
and when $|\psi_{a,b}\rangle$ is chosen from the intersection the range of $P_{a,b}$ and ${\cal H}^{\otimes m}$,
the state $W_m |\phi\rangle$ belongs to ${\cal H}^{\otimes m}$.
The estimator $(W_m|\phi,\{W_m M_{\hat\theta}W_m^\dagger \})$ simulates the estimator $(|\phi\rangle,\{M_{\hat\theta}\})$.
So, 
the optimization of estimator on the single input system $L^2(\real^2)$ 
with constraint concerning maximum of photon number
is mathematically equivalent with that on the $m$-fold tensor product system ${\cal H}$.

Overall, multiple application setting is a physical situation different from single application setting.
This is because multiple application requires multiple application on the same system $L^2(\real^2)$.
However, this setting can be mathematically simulated by single application setting.
This kind of simulation plays an important role in the next section.

\section{Locally asymptotic mini-max criterion}\Label{s4}
In this section, we consider the locally asymptotic mini-max criterion.
In statistics, we often treat mini-max criterion, in which, we optimize the worst case.
In the following, we optimize the worst MSE among the $\epsilon$-neighborhood of the given point $\theta_0$.
We treat the asymptotic behavior of this optimum value and consider the limit $\epsilon \to 0$ concerning 
the asymptotic coefficient. That is, we define
\begin{align*}
 C_{av,\theta}  
:=  &
\lim_{\epsilon \to 0} \lim_{E \to \infty} 
E^2 C_{av,\theta} (E,\epsilon) \\
 C_{av,\theta} (E,\epsilon) 
:=& 
\inf_{|\phi\rangle}
\{ C_{\theta} (|\phi\rangle :\epsilon)
| {(N^2)}_{av}(|\phi\rangle) \le E^2 \} \\
 C_{\theta_0} (|\phi\rangle :\epsilon) 
:=& 
\inf_{M}
\sup_{\theta\in U(\epsilon,\theta_0)} E_{\theta} (|\phi\rangle,M)
\\
 C_{max,\theta} 
 :=&
\lim_{\epsilon \to 0} \lim_{E \to \infty} 
E^2 C_{max,\theta}(E,\epsilon) \\
 C_{max,\theta}(E,\epsilon)
:= & \inf_{|\phi\rangle}
\{ C_{\theta}(|\phi\rangle :\epsilon)
|{N}_{max}(|\phi\rangle) \le E
 \} .
\end{align*}
We also define similar values with multiple application.
\begin{align*}
&C_{av,\theta_0,m}(E,\epsilon) \\
:=&
\inf_{(|\phi\rangle,M)}
\{ \sup_{\theta\in U(\epsilon,\theta_0)} E_{\theta} (|\phi\rangle,M)
| {(N_m^2)}_{av}(|\phi\rangle) \le E^2 m^2 \} \\
&C_{max,\theta_0,m} \\
:=&
\inf_{(|\phi\rangle,M)}
\{ \sup_{\theta\in U(\epsilon,\theta_0)} E_{\theta} (|\phi\rangle,M)
|~|\phi\rangle \in {\cal H}^{\otimes m} \} ,
\end{align*}
where 
$(|\phi\rangle,M)$ is an estimator on $L^2(\real^2)^{\otimes m}$.
As is discussed in  Section \ref{s3},
estimators on $L^2(\real^2)^{\otimes m}$
can be simulated by estimators on $L^2(\real^2)$ with large photon number states.
Hence,
\begin{align*}
C_{av,\theta} =&
\lim_{\epsilon \to 0} \lim_{m \to \infty} 
m^2 E^2 C_{av,\theta,m}(1,\epsilon) \\
C_{max,\theta} =&
\lim_{\epsilon \to 0} \lim_{m \to \infty} m^2 C_{max,\theta,m} .
\end{align*}
Further, Proposition 1 of \cite{Ha5} implies that
\begin{align*}
\lim_{\epsilon \to 0} \lim_{m \to \infty} m^2 C_{av,\theta,m}(1,\epsilon) 
&\le \tilde{C}_{av,\theta} \\
\lim_{\epsilon \to 0} \lim_{m \to \infty} m^2 C_{max,\theta,m} 
&\le
\tilde{C}_{max,\theta}.
\end{align*}
That is,
\begin{align*}
C_{av,\theta} &\le \tilde{C}_{av,\theta} \\
C_{max,\theta} &\le \tilde{C}_{max,\theta}.
\end{align*}

\section{Global mini-max criterion}\Label{s5}
Next, we consider global optimization, i.e., global mini-max criterion.
For this purpose, we consider group covariant condition for our POVM $M$.
In this optimization with photon number constraint,
we can restrict the input state on the subspace 
${\cal H}_1$ spanned by $\{|n,0\rangle\} \cup \{|0,n\rangle\} $.
A POVM $M=\{M_{\hat\theta}\}$ is covariant when 
there exists a vector $|T\rangle$ such that
$M_{\hat\theta}d\hat\theta=\frac{1}{2\pi} U_{\hat{\theta}}|T\rangle\langle T|U_{\hat{\theta}}^\dagger d\hat\theta$,
where
$|T\rangle =\sum_{n=0}^{\infty}e^{i\theta_n}|n,0\rangle+
\sum_{n=1}^{\infty}e^{i\theta_n'}|0,n\rangle$,
and $\theta_n$ and $\theta_n$ are arbitrary coefficients.
As is shown by Holevo \cite{HoC},
since the stochastic behavior of the error $\hat\theta-\theta$
of the covariant POVM does not depend on the true parameter $\theta$,
any input state $|\phi\rangle$ satisfies
\begin{align*}
\inf_{M} 
\max_\theta E_\theta(|\phi\rangle,M)
=
\inf_{M:\hbox{cov}}\max_\theta E_\theta(|\phi\rangle,M).
\end{align*}
Hence, when we optimize the worst value 
$\max_\theta E_\theta(|\phi\rangle,M)$,
it is sufficient to treat the covariant POVM.
That is, our estimator is written by the pair of input state 
$|\phi\rangle= 
\sum_{n=0}^{\infty}a_n|n,0\rangle+\sum_{n=1}^{\infty}a_{-n}|0,n\rangle$ 
and $|T\rangle=\sum_{n=0}^{\infty}e^{i\theta_n}|n,0\rangle+
\sum_{n=1}^{\infty}e^{i\theta_n'}|0,n\rangle$,
where $\vec{a}=
\{a_n\}_{n=-\infty}^{\infty}\in L^2(\integer)$.
This estimator is simulated by the pair of
the input state 
$|\phi'\rangle= 
\sum_{n=0}^{\infty} e^{-i\theta_n} a_n|n,0\rangle+\sum_{n=1}^{\infty}e^{-i\theta_n'}a_{-n}|0,n\rangle$ 
and the covariant POVM given by 
$|T_0\rangle=\sum_{n=0}^{\infty}|n,0\rangle+
\sum_{n=1}^{\infty}|0,n\rangle$.
This is because 
$|\langle T|U_{\hat\theta}^\dagger U_{\theta}|\phi\rangle|^2
=|\langle T_0|U_{\hat\theta}^\dagger U_{\theta}|\phi'\rangle|^2$.
Hence, it is sufficient to treat the case with $|T\rangle=|T_0\rangle$.
In this case, we define 
$a_n':=e^{-i\theta_n} a_n$ for $n \ge 0$
and $a_{-n}':=e^{-i\theta_n'} a_{-n}$ for $n > 0$.
Then,
when the error $\hat\theta-\theta$ obeys the square 
$|{\cal F}^{-1}(\vec{a}')(\theta)|^2$
of the inverse Fourier transform 
${\cal F}^{-1}(\vec{a}')(\theta):=\sum_{n=-\infty}^{\infty} a_n' e^{-i n \theta}$
of discrete series of $\vec{a}'=\{a_n'\}_{n=-\infty}^{\infty}$.

Thus, we obtain 
\begin{align*}
C(|\phi\rangle) 
:=&
\inf_{M} \sup_\theta E_\theta(|\phi\rangle,M) \\
=&
\inf_{\vec{a'}:|a_n'|=|a_n|}
\{\frac{1}{2\pi}\int_{-\pi}^\pi
\hat{\theta}^2
|{\cal F}^{-1}(\vec{a'})(\hat\theta)|^2
d\hat{\theta} 
\end{align*}
and
\begin{align}
&C_{av}(E) \nonumber\\
:=&
\inf_{(|\phi,M)} 
\{\sup_\theta E_\theta(|\phi\rangle,M) |
\langle \phi| \hat{N}^2 |\phi\rangle \le E^2
 \}\nonumber \\
=&
\inf_{\vec{a}:\|\vec{a}\|=1}
\{\frac{1}{2\pi}\int_{-\pi}^\pi
\hat{\theta}^2
|{\cal F}^{-1}(\vec{a})(\hat\theta)|^2
d\hat{\theta} 
|
{(N^2)}_{av}(\vec{a}) \le E^2
\} \Label{10-27-1}\\
&C_{max}(E) \nonumber \\
:=&
\inf_{(|\phi,M)} 
\{\sup_\theta E_\theta(|\phi\rangle,M) |
\langle \phi| \hat{N}^2 |\phi\rangle \le E^2
 \} \nonumber \\
= &
\inf_{\vec{a}:\|\vec{a}\|=1}
\{
\frac{1}{2\pi}\int_{-\pi}^\pi
\hat{\theta}^2
|{\cal F}^{-1}(\vec{a})(\hat\theta)|^2
d\hat{\theta} 
|
{N}_{max}(\vec{a}) \le E \} \Label{10-27-2},
\end{align}
where 
${(N^2)}_{av}(\vec{a}):= \sum_{n} n^2|a_n |^2 $,
${N}_{max}(\vec{a}):= \max_{n}\{|n| ~| a_n\neq 0\}$,
and $\|\vec{a}\|^2:=\sum_{n=-\infty}^{\infty}|a_n|^2$.

From the definitions, we obtain
\begin{align}
C(|\phi\rangle) & \ge C_{\theta}(|\phi\rangle: \epsilon) \Label{12-28-2} \\
C_{av}(E) & \ge C_{av,\theta}(E,\epsilon) \Label{10-27-5}\\
C_{max}(E) & \ge C_{max,\theta}(E,\epsilon)\Label{10-27-6},
\end{align}
and the asymptotic limits are defined by
\begin{align*}
C_{av} &:= \lim_{E \to \infty} E^2 C_{av}(E) \\
C_{max}&:= \lim_{E \to \infty} E^2 C_{max}(E) .
\end{align*}

Given an arbitrary sequence $\vec{a_{E}=\{a_{E,n}}\}_{n=-\infty}^{\infty}$ of elements of $L^2(\integer)$,
we focus on the function 
\begin{align}
f(x):= \lim_{E \to \infty} \sqrt{E} a_{E,Ex}.\Label{10-27-10}
\end{align}
Then, the relations (\ref{10-27-1}) and (\ref{10-27-2}) imply that
\begin{align}
C_{av}
&=\inf_{f \in L^2(\real):\|f\|=1}
\{
\langle f|P^2|f\rangle | \langle f|Q^2|f\rangle \le 1 \}\Label{10-27-3} \\
C_{max}
&=\inf_{f \in L^2([-1,1]):\|f\|=1}
\langle f|P^2|f\rangle ,\Label{10-27-4}
\end{align}
where $Q$ and $P$ are the position and momentum operators on $L^2(\real)$.
The equation (\ref{10-27-4}) is the same as the result obtained by Imai and Hayashi \cite{IH}.

Next, we focus on the specific sequence of the input states $|\phi_E\rangle$
whose coefficient satisfies (\ref{10-27-10}).
Similar to (\ref{10-27-3}) and (\ref{10-27-4}), we obtain
\begin{align}
& \lim_{E \to \infty} E^n C(|\phi_E\rangle) \nonumber\\
=&
\inf_{g \in L^2(\real): |g(x)|=|f(x)|} \langle g|P^2|g\rangle.
\Label{12-28-1}
\end{align}

The uncertainly relation implies that
\begin{align*}
C_{av}=\frac{1}{4}
\end{align*}
and the infimum is attained when $f(x)=e^{-\frac{x^2}{4}}$.
The latter is given by
\begin{align*}
C_{max}=\frac{\pi^2}{4}
\end{align*}
and the infimum is attained when $f(x)=(2\pi)^{1/4}\sin \frac{\pi(1+x)}{2}$.
Comparing the right hand sides of (\ref{10-27-1}) and (\ref{10-27-2}) with those of (\ref{10-27-3}) and (\ref{10-27-4}),
the bounds $C_{av}$ and $C_{max}$ can be attained by 
sequence of input states
\begin{align}
c_1(\sum_{n=0}^{\infty} e^{-\frac{n^2}{4 E^2}} |n,0\rangle+\sum_{n=1}^{\infty} e^{-\frac{n^2}{4 E^2}} |0,n\rangle) 
\Label{10-27-20}
\end{align}
and
\begin{align}
&c_2(\sum_{n=0}^{E} \sin\frac{\pi(E+n+1)}{2E+2} |n,0\rangle \nonumber \\
&\quad +\sum_{n=1}^{E} \sin\frac{\pi(E+n+1)}{2E+2} |0,n\rangle),
\Label{10-27-21}
\end{align}
where $c_1$ and $c_2$ are normalized constants.

From the relations (\ref{10-27-5}) and (\ref{10-27-6}), we obtain
\begin{align}
C_{av} & \ge C_{av,\theta}\Label{10-27-7} \\
C_{max} & \ge C_{max,\theta}.\Label{10-27-8}
\end{align}
As is shown in Proposition 3 of Hayashi \cite{Ha5},
the bounds 
$ C_{av,\theta} =\lim_{\epsilon \to 0} \lim_{m \to \infty} m^2 E^2 C_{av,\theta,m}(\epsilon,E) $
and 
$C_{max,\theta} =\lim_{\epsilon \to 0} \lim_{m \to \infty} m^2 C_{max,\theta,m}$
can be globally attained by two-step methods \cite{4,5} asymptotically.
Since any estimator on the multiple system $L^2(\real^2)^{\otimes m}$ can be simulated by a suitable estimator on the single input system $L^2(\real^2)$,
the opposite inequalities of (\ref{10-27-7}) and (\ref{10-27-8}) hold.
That is, combining (\ref{12-28-6}) and (\ref{12-28-7}), we obtain
\begin{align}
C_{av,\theta} & =C_{av}  =\frac{1}{4}= \tilde{C}_{av,\theta} \\ 
C_{max,\theta} & =C_{max}  =\frac{\pi^2}{4} > \frac{1}{4}=\tilde{C}_{max,\theta}.
\end{align}
So, when we adopt the maximum photon number constraint,
there is a non-negligible difference between the local asymptotic mini-max bound 
and the bound with locally unbiased condition.

\section{Analysis on noon state}\Label{s6}
Now, we consider the asymptotic limit of the performance with the locally asymptotic min-max criterion
when the noon state $\frac{1}{\sqrt{2}}(|n,0\rangle +|0,n\rangle)$ is inputed and the covariant measurement is applied.
In this case, first, we fix an arbitrary small real number $\epsilon >0$
and focus on the neighborhood $U(\epsilon,\theta_0)$ for an arbitrary point $\theta_0$.
When $n > 2\pi/\epsilon$,
the relation 
$U_\theta \frac{1}{\sqrt{2}}(|n,0\rangle +|0,n\rangle)=
U_{\theta+2\pi/n } \frac{1}{\sqrt{2}}(|n,0\rangle +|0,n\rangle)$
holds and 
$\theta$ and $\theta + \frac{2\pi}{n}$ belongs to $U(\epsilon,\theta_0)$.
That is, we cannot distinguish two parameters $\theta$ and $\theta + \frac{2\pi}{n}$.
Even if we could estimate the unknown parameter $\theta$ mod $ \frac{2\pi}{n}$ perfectly, 
it is not easy to estimate the parameter $\theta$ in the parameter space $U(\epsilon,\theta_0)$.
In the following, we consider the case of $\theta=\theta_0 -\epsilon$ mod $\frac{2\pi}{n}$.
Assume that we could the unknown parameter $\theta$ mod $\frac{2\pi}{n}$.
In this case, there are still $K:=\lfloor \frac{n\epsilon}{\pi} \rfloor$
candidates with the width $\frac{2 \pi}{n}$ in the parameter space $U(\epsilon,\theta_0)$.
Assume that, we decide the true parameter to be $\frac{2 \pi}{n}i+\theta_0-\epsilon$ with the probability $p_i$.
Then, the mini-max error is evaluated as follows.
\begin{align*}
& \max_{j=1, \ldots, K}
\sum_{i=1}^K p_i (\frac{1}{2 n\pi})^2(j-i)^2\\
=&
(\frac{2\pi}{n})^2
\max_{j=1, \ldots, K}
(j^2 -2 j \sum_{i=1}^K i p_i + \sum_{i=1}^K i^2 p_i) \\
= &
(\frac{2\pi}{n})^2
(\max_{j=1, \ldots, K}
(j - \sum_{i=1}^K i p_i )^2 + \sum_{i=1}^K i^2 p_i-( \sum_{i=1}^K i p_i)^2) \\
\ge &
(\frac{2\pi}{n})^2
\max_{j=1, \ldots, K}
(j - \sum_{i=1}^K i p_i )^2 \\
\ge & 
(\frac{2\pi}{n})^2 (K/2)^2
=
(\frac{\pi}{n} \lfloor \frac{n\epsilon}{\pi} \rfloor )^2 .
\end{align*}
That is, we obtain  
\begin{align*}
C_{\theta_0} (\frac{1}{\sqrt{2}}(|n,0\rangle +|0,n\rangle) :\epsilon) 
\ge
(\frac{\pi}{n} \lfloor \frac{n\epsilon}{\pi} \rfloor )^2.
\end{align*}
The lower bound $(\frac{\pi}{n} \lfloor \frac{n\epsilon}{\pi} \rfloor )^2 $
converges $\epsilon^2$. 
This means that the mini-max of the mean square error does not decrease when $n$ increase.
Hence,  
\begin{align}
\lim_{n \to \infty} n^2 C_{\theta_0} (\frac{1}{\sqrt{2}}(|n,0\rangle +|0,n\rangle) :\epsilon) 
=\infty.
\Label{12-29-1}
\end{align}

Next, we consider the asymptotic limit of the performance with the global asymptotic min-max criterion
when the noon state $\frac{1}{\sqrt{2}}(|n,0\rangle +|0,n\rangle)$.
By using (\ref{12-29-1}) and (\ref{12-28-2}), the relation 
\begin{align}
\lim_{n \to \infty} n^2 C (\frac{1}{\sqrt{2}}(|n,0\rangle +|0,n\rangle) ) =\infty
\Label{12-28-4}
\end{align}
 holds.
This fact can be shown in another method based on (\ref{12-28-1}).
In fact, 
the $L^2$ function corresponding to the noon state via (\ref{10-27-10}) is 
$f_0(x):=\frac{1}{\sqrt{2}}(\delta(x-1)+\delta(x+1))$.
The function 
$f_\eta (x):=\frac{1}{\sqrt{2}}(\delta(x-1)+e^{i\eta} \delta(x+1))$
satisfies $\langle f_\eta | P^2|f_\eta\rangle=\infty$.
Thus, the relation (\ref{12-28-1}) yields (\ref{12-28-4}).

Next, we replace the number state by the coherent state.
In the following, the coherent state with complex amplitude $\alpha_1,\alpha_2$ is written as
$|\alpha_1,\alpha_2\rangle_c$.
Then, we consider the case when 
the covariant measurement is applied and we input
the noon state $\frac{1}{\sqrt{2}}(|n,0\rangle_c +|0,n\rangle_c)$,
which is called the coherent noon state.
The $L^2$ function corresponding to the coherent noon state via (\ref{10-27-10}) is 
the same function $f_0$.
Hence, the relation (\ref{12-28-1}) yields 
\begin{align}
\lim_{n \to \infty} n^2 C (\frac{1}{\sqrt{2}}(|n,0\rangle_c +|0,n\rangle_c) ) =\infty.
\end{align}

\section{Conclusion}\Label{s7}
We have discussed three kinds of bounds
of MSE in the phase estimation with photon number constraints.
The first is the global minimax bound.
The second is the local minimax bound.
The third is the bound with the locally unbiased condition.
These bounds have been treated with two kinds of constraints:
One is 
the constraint for the average of the square of the photon number concerning the input state.
The other is 
the constraint for the maximum photon number concerning the input state.
We have shown that the asymptotic limits 
$C_{av}$, $C_{av,\theta}$, and $\tilde{C}_{av,\theta}$ of three kinds of bounds coincide
under the first constraint.
However, these bounds 
$C_{max}$, $C_{max,\theta}$, and $\tilde{C}_{max,\theta}$ 
do not coincide under the second kind of constraint.
In fact, the locally unbiased condition and Fisher information are originally 
mathematical concepts.
In the independent and identical distributed setting of state estimation, 
these values provide the asymptotic bound of MSE\cite{4,5,F1}.
However, these values do not provide the operational meaning in general.
That is, these values have no operational meaning when they do not coincide with operational values.
Hence, 
under the second kind of constraint, only these bounds $C_{max}$ and $C_{max,\theta}$ have the real meaning.

Further, we should be careful of the meaning of the noon state and the coherent noon state.
When these states are inputed, the optimal SLD Fisher information is realized.
However, under the local mini-max criterion, as is discussed in section 6,
the MSE of the noon input state does not convergence to zero, 
and the noon input state is far from the optimal input.
The optimal input is given in (\ref{10-27-20}) and (\ref{10-27-21}).
Hence, in order to realize the high performance phase estimation, 
it is desired to implement the input states (\ref{10-27-20}) and (\ref{10-27-21}).

\noindent{\sf Acknowledgement}\quad 

The author was partially supported by a Grant-in-Aid for Scientific Research
in the Priority Area `Deepening and Expansion of Statistical 
Mechanical Informatics (DEX-SMI)', No. 18079014
and a MEXT Grant-in-Aid for Young Scientists (A) No. 20686026.
The Centre for Quantum Technologies is funded by the Singapore Ministry of Education
and the National Research Foundation as part of the Research Centres of Excellence
programme.
The author thanks Dr. William J. Munro and Mr. Akihito Soeda for helpful comments.
He also thanks the referees for helpful comments
concerning this manuscript.


\end{document}